\documentclass[aps,pre,twocolumn,superscriptaddress,longbibliography]{revtex4}

\usepackage{amsmath}
\usepackage{amssymb}
\usepackage{wrapfig}
\usepackage{mathrsfs}
\usepackage{cases}
\usepackage{braket}
\usepackage{verbatim}
\usepackage{latexsym}
\usepackage{multirow}
\usepackage{color}
\usepackage{graphicx}
\usepackage{url}
\usepackage[toc, page]{appendix}
\usepackage[T1]{fontenc}
\usepackage[
	colorlinks=true,
	linkcolor=blue,
	citecolor=blue,
	urlcolor=blue,
	hyperindex,
	breaklinks
]{hyperref}

\def\v#1{\mbox{\boldmath $#1$}}

\def\pd#1#2{\frac{\partial #1}{\partial #2}}

\begin{document}

\title{Retrieving the ground state of spin glasses using thermal noise:\\
Performance of quantum annealing at finite temperatures}

\author{Kohji Nishimura}
\affiliation{Department of Physics, Tokyo Institute of Technology, Oh-okayama, Meguro-ku, Tokyo 152-8551, Japan}

\author{Hidetoshi Nishimori}
\affiliation{Department of Physics, Tokyo Institute of Technology, Oh-okayama, Meguro-ku, Tokyo 152-8551, Japan}

\author{Andrew J. Ochoa}
\affiliation{Department of Physics and Astronomy, Texas A$\&M$ University,
College Station, Texas 77843-4242, USA}

\author{Helmut G. Katzgraber}
\affiliation{Department of Physics and Astronomy, Texas A$\&M$ University,
College Station, Texas 77843-4242, USA}
\affiliation{Santa Fe Institute, 1399 Hyde Park Road, Santa Fe, New Mexico
87501, USA}

\date{\today}

\begin{abstract}
	We study the problem to infer the ground state of a spin-glass Hamiltonian using data from another Hamiltonian with interactions disturbed by noise from the original Hamiltonian, motivated by the ground-state inference in quantum annealing on a noisy device. It is shown that the average Hamming distance between the inferred spin configuration and the true ground state is minimized when the temperature of the noisy system is kept at a finite value, and not at zero temperature. We present a spin-glass generalization of a well-established result that the ground state of a purely ferromagnetic Hamiltonian is best inferred at a finite temperature in the sense of smallest Hamming distance when the original ferromagnetic interactions are disturbed by noise.  We use the numerical transfer-matrix method to establish the existence of an optimal finite temperature in one- and two-dimensional systems. Our numerical results are supported by mean-field calculations, which give an explicit expression of the optimal temperature to infer the spin-glass ground state as a function of variances of the distributions of the original interactions and the noise.  The mean-field prediction is in qualitative agreement with numerical data. Implications on postprocessing of quantum annealing on a noisy device are discussed.
\end{abstract}


\maketitle

\section{Introduction}

Quantum annealing \cite{Kadowaki1998, KadowakiThesis, Finnila1994, Arnab2008, Santoro2006, Morita2008, Bapst2013, Farhi2001} is the quantum-mechanical counterpart of simulated annealing \cite{SA}, an optimization method used to find the ground state of, e.g., an Ising model. Because combinatorial optimization problems can be formulated as the ground-state search of an Ising model \cite{lucas2014}, typically with spin-glass-like complex interactions, the development of efficient methods to solve these computationally hard problems is an important target of current research activities. Quantum annealing has attracted a great deal of attention in recent years partly due to the introduction of its hardware implementation, the D-Wave quantum annealer \cite{Johnson2011}.  Evidence has been provided that the machine indeed runs quantum mechanically \cite{Boixo2013,Lanting2014,Albash2015,Boixo2016,Johnson2011,Dickson2013,Boixo2014,Albash2015v2,Crowley2014,katzgraber:15,mandra:16}; however, there remain open questions before the device becomes practically useful. One of the problems is the control error, i.e., imperfections in the setting of parameter values of the Ising Hamiltonian in the device \cite{Perdomo-Ortiz2015,zhu:16}.  Because it is difficult to set the parameter values (interactions and local fields of the Hamiltonian) with a high precision, the device might be attempting to find the ground state of the wrong Hamiltonian, thus compromising the reliability of the final output. This phenomenon arises in any analog device like the D-Wave quantum annealer and it is crucial to devise and implement ingenious methods to mitigate the influence of control errors. One approach is quantum error correction \cite{lidar:08,lidar:13,pudenz:13,pudenz:15,vinci:15}, however, at the cost of decreasing the number of available logical qubits.

A closely related problem has been analyzed in the context of classical error-correcting codes, in which a message is encoded and then transmitted through a noisy channel. The receiver has to correct errors in the received noisy signal and retrieve the original message.  It is known that a certain type of error-correcting codes can be formulated in terms of the theory of spin-glasses \cite{Sourlas,nsmrbook}. In this formulation, the task to retrieve the original message is translated to the inference of the ground state of an Ising model with uniform ferromagnetic interactions using only the information of a spin-glass Hamiltonian derived from the original ferromagnetic Hamiltonian by the application of noise to the interactions. It has been shown numerically \cite{rujan1993}, as well as analytically \cite{nishimori1993,sourlas1994,iba1999}, that the best performance to retrieve the ferromagnetic ground state from the spin-glass Hamiltonian is achieved at a finite temperature rather than at zero temperature, since the Hamming distance of the retrieved spin configuration to the true ground state is a nonmonotonic function of the temperature. In fact, recent experiments on the D-Wave quantum annealer \cite{chancellor:16} study the trivial ferromagnetic case and illustrate that decoding is more efficient at finite temperature.

In the present paper we generalize the above formulation for error-correcting codes to the situation where the original (noiseless) Hamiltonian already has randomness in the interactions, i.e., a spin-glass Hamiltonian.  The random interactions are then disturbed by (Gaussian) noise, and the task is to find a spin configuration closest to the ground state of the original Hamiltonian using only the Hamiltonian with disturbed interactions. One of the important motivations to study such a problem lies in the noise-mitigation task of analog devices as explained above: One is faced with the problem to infer the correct ground state of a Hamiltonian with random interactions out of data produced from the Hamiltonian with noise in addition to the original random interactions.  Because it is very difficult to develop a general theory for this situation as was done for error-correcting codes \cite{nishimori1993,nsmrbook}, we use numerical methods for one- and two-dimensional models supplemented by a mean-field-type approach. Our result shows clearly that the original ground state is better inferred at finite temperature, rather than at zero temperature. Therefore, tuning the D-Wave quantum annealer to the optimal decoding temperature might actually assist in mitigating the effects of analog noise in the device.

The paper is organized is as follows. In Sec.~\ref{sec:Formulation}, we formulate the problem. The results of numerical calculations are described in Sec.~\ref{sec:NumAnalysis}.  An analysis using mean-field theory is given in Sec.~\ref{sec:MFA}, followed by concluding remarks in Sec.~\ref{sec:Conclusion}. Technical details are delegated to the Appendix.

\section{Formulation of the problem} 
\label{sec:Formulation}

The goal is to infer the original ground state of a Hamiltonian whose interactions $J_{i_1, \ldots ,i_p}$ follow a Gaussian distribution and are disturbed by noise, $J_{i_1, \ldots ,i_p} \to \tilde{J}_{i_1, \ldots ,i_p}$.
We write the original spin-glass Hamiltonian with general many-body interactions as,
\begin{align}
	\label{eq:2:Hsi}
	H(\v{S}) = -\sum_{i_1, \ldots ,i_p} J_{i_1, \ldots ,i_p} S_{i_1} \cdots S_{i_p},
\end{align}
where $\v{S}(=S_1, S_2, \ldots, S_N)$ is the set of Ising spins, i.e.,
$S_i \in \{\pm 1\}$. The original interactions $J_{i_1, \ldots ,i_p}$ are generated from a Gaussian distribution $P(J_{i_1, \ldots ,i_p})$ with mean unity and variance $\sigma^2$. The ground state of this Hamiltonian is given by the sign of the zero-temperature limit of the thermal expectation value, i.e., 
\begin{align}
	\mathcal{S}^{(0)}_{i} &= \lim_{\beta_0 \to \infty}\mathrm{sgn}\Braket{S_i}_{\beta_0} \nonumber \\
	\label{eq:2:Sground}
	&= \lim_{\beta_0 \to \infty}\mathrm{sgn}\left(\frac{\mathrm{Tr}_{{\scriptsize \v{S}}} \, S_i \exp{[-\beta_0 H(\v{S})]}}{\mathrm{Tr}_{{\scriptsize \v{S}}} \exp{[-\beta_0 H(\v{S})]}}\right) .
\end{align}
We next introduce disturbed interactions $\tilde{J}_{i_1, \ldots ,i_p}$ by adding a noise term $\xi_{i_1, \ldots ,i_p}$ to the original interactions,
\begin{align}
    \label{eq:2:addnoise}
    \tilde{J}_{i_1, \ldots ,i_p} \equiv J_{i_1, \ldots ,i_p}+\xi_{i_1, \ldots ,i_p} .
\end{align}
The variables $\xi_{i_1, \ldots ,i_p}$ are distributed according to a  Gaussian distribution $P(\xi_{i_1, \ldots ,i_p})$ with zero mean and variance $\gamma^2$. The problem is to find a spin configuration closest to the ground state of the {\em original} Hamiltonian Eq. (\ref{eq:2:Hsi}) from the {\em noisy} Hamiltonian, Eq. (\ref{eq:2:Htilde}) below, at finite temperature $T$ or the inverse temperature $\beta =1/T$,
\begin{align}
	\label{eq:2:ftd}
	\mathcal{S}^{(\beta)}_{i} &= \mathrm{sgn}\Braket{S_i}_{\beta} = \mathrm{sgn}\left(\frac{\mathrm{Tr}_{{\scriptsize \v{S}}}\, S_i \exp{[-\beta \tilde{H}(\v{S})]}}{\mathrm{Tr}_{{\scriptsize \v{S}}} \exp{[-\beta \tilde{H}(\v{S})]}}\right) ,
\end{align}
where
\begin{align}\label{eq:2:Htilde}
	\tilde{H}(\v{S}) = -\sum_{i_1, \ldots ,i_p} \tilde{J}_{i_1, \ldots ,i_p}S_{i_1} \cdots S_{i_p}
\end{align}
represents the Hamiltonian with the added noise.  Notice that we are interested only in the {\em sign} of the spin average, not the magnitude, as indicated in Eq.~(\ref{eq:2:ftd}). As a measure of similarity between the ground state of the original Hamiltonian and the spin configuration of the noisy system, we define the overlap $M(T)$ as
\begin{align}
	\label{eq:2:Mb}
	M(T) = \int \prod d{J}P(J)\int \prod d{\xi}P(\xi)\, \mathcal{S}^{(0)}_{i} \mathcal{S}^{(\beta)}_{i} , 
\end{align}
where the products run over the set of interactions. The overlap $M(T)$ is closely related to the average Hamming distance $D(T)$,
\begin{equation}
    D(T)=\int \prod d{J}P(J)\int \prod d{\xi}P(\xi)\, \sum_{i=1}^N\frac{(\mathcal{S}^{(0)}_{i}-\mathcal{S}^{(\beta)}_{i})^2}{4}
\end{equation}
by the relation
\begin{equation}
    D(T)=\frac{N}{2}\big(1-M(T)\big).
\end{equation}
A large overlap $M(T)$ means a small Hamming distance. The Hamming distance is a standard measure of the quality of error-correcting codes \cite{nsmrbook}, and we adopt it as the quantity to be minimized [or the overlap $M(T)$ to be maximized] in the present paper.

Optimal decoding occurs at the point where $M(T)$ has a maximum. Here we analyze how this maximum depends on the strength of the disorder $\gamma$.

When $\sigma = 0$, the Hamiltonian in Eq.~(\ref{eq:2:Hsi}) represents the ferromagnetic Ising model. Then, $\mathcal{S}^{(0)}_{i}=1$ for all $i$ (or maybe $\mathcal{S}^{(0)}_{i}=-1~\forall i$), and the overlap $M(T)$ is identical to that of error-correcting codes with all original bits being unity \cite{nsmrbook}.  It is known in this case that the overlap in Eq.~(\ref{eq:2:Mb}) takes a maximum value at  $T_{\mathrm{N}} = \gamma^2$, the so-called Nishimori temperature \cite{rujan1993,nishimori1993,sourlas1994,iba1999,nsmrbook}.
It is difficult to apply the same theory to the case where $\sigma \neq 0$ due to the lack of proper symmetry. We therefore use numerical methods in the following section to study the behavior of the disordered system.

\section{Numerical analysis} 
\label{sec:NumAnalysis}

\begin{figure}
	\centering
	\includegraphics[width=\linewidth]{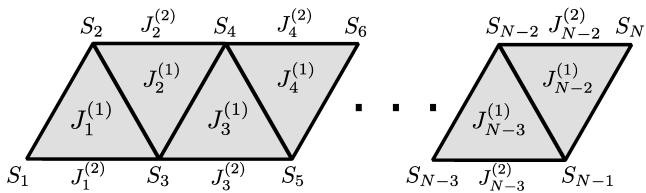}
	\caption{Ising spin glass on the triangular ladder.}
	\label{fig:2:ladder}
\end{figure}

\begin{figure}
	\centering
	\includegraphics[width=\linewidth]{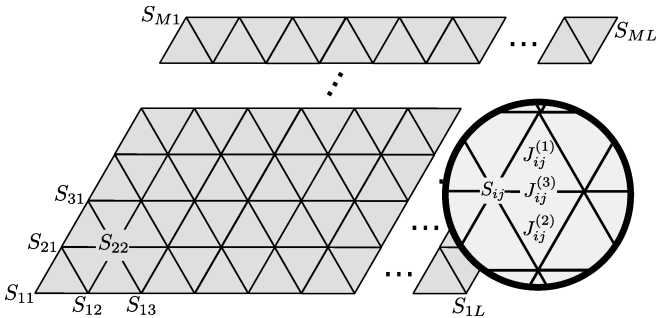}
	\caption{Ising spin glass on a stacked triangular ladder.}
	\label{fig:2:2dmodel}
\end{figure}

Following Ref.~\cite{rujan1993}, we apply the numerical transfer-matrix method to a triangular ladder with two- and three-body interactions, as depicted in Fig.~\ref{fig:2:ladder}, using free boundary conditions,
\begin{align}
    \label{eq:2:ladder}
	H = -\sum_{i=1}^{N-2}\left(J^{(1)}_i S_i S_{i+1} S_{i+2}+J^{(2)}_i S_i S_{i+2}\right) .
\end{align}
Furthermore, we study a quasi-two-dimensional system with triangular ladders stacked on top of each other as depicted in Fig.~\ref{fig:2:2dmodel} using the transfer-matrix method.


The overlap for finite-size systems
		\begin{align}
			M(T) = \frac{1}{N}\sum_{i=1}^{N} \mathcal{S}^{(0)}_{i}\mathcal{S}^{(\beta)}_{i}
		\end{align}
is calculated for these lattices with the ground-state configuration $\mathcal{S}^{(0)}_{i}$ determined by the Viterbi algorithm \cite{viterbi1967}, which is a zero-temperature transfer-matrix method.  We take the configurational average over the distributions of the original interactions and noise by sampling $400$ disorder realizations.
The number of spins is $N=10^5$ for the ladder and $N=LM$ with $L=10^3$ and $M=10$ for stacked ladders. The standard deviations of the original interactions are chosen to be $\sigma = 1.0$, $1.4$, $2.0$, and $3.0$.  For each value of $\sigma$, we apply noise with a typical strength of $\gamma = 0.1$, $0.2$, $0.25$, $0.3$, $0.4$, $0.5$, $0.6$, $0.7$, $0.8$, $0.9$, $1.0$, $1.25$, $1.5$, and $2.0$, respectively.

\begin{figure*}
			\includegraphics[width=\columnwidth]{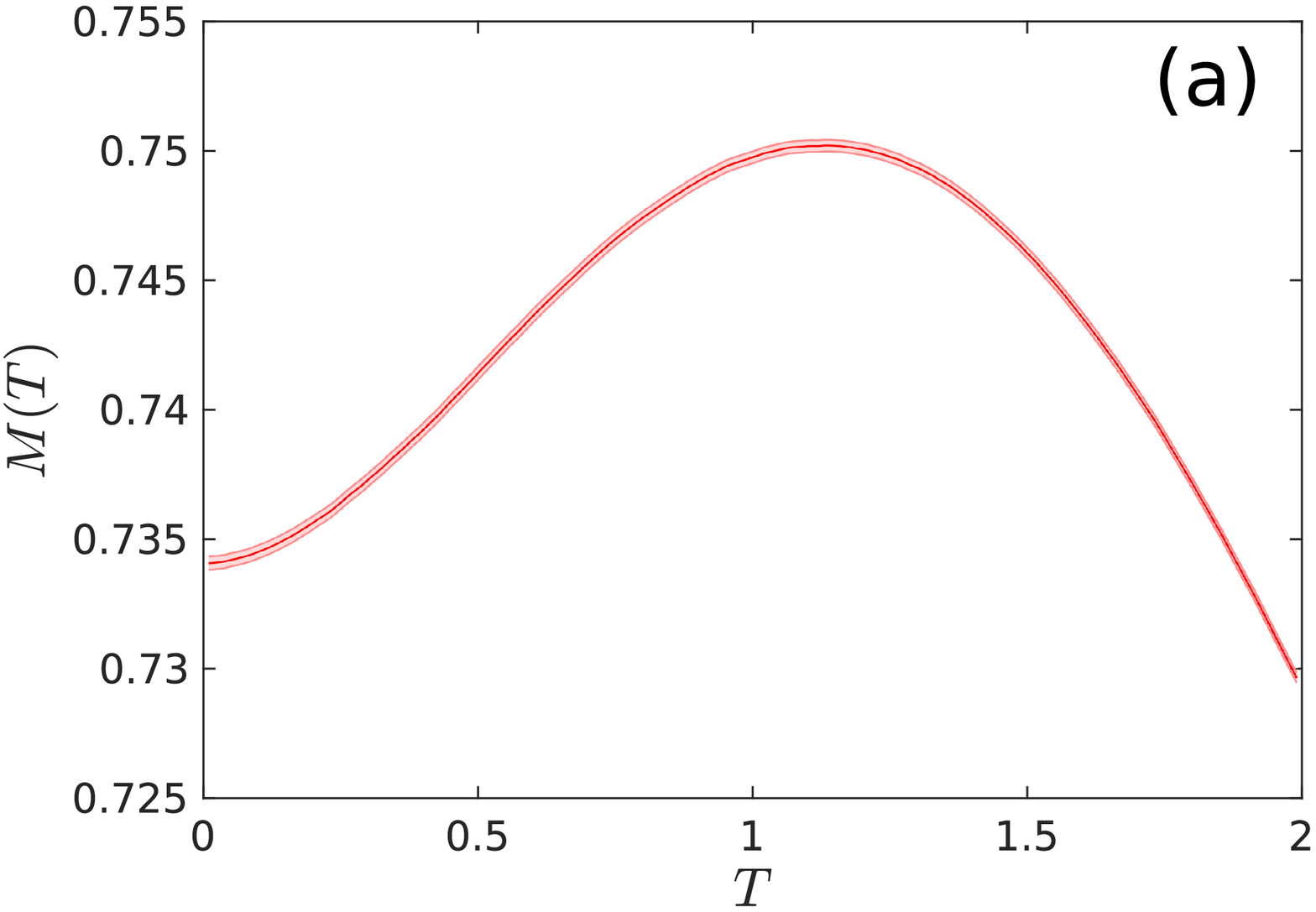}
			\includegraphics[width=\columnwidth]{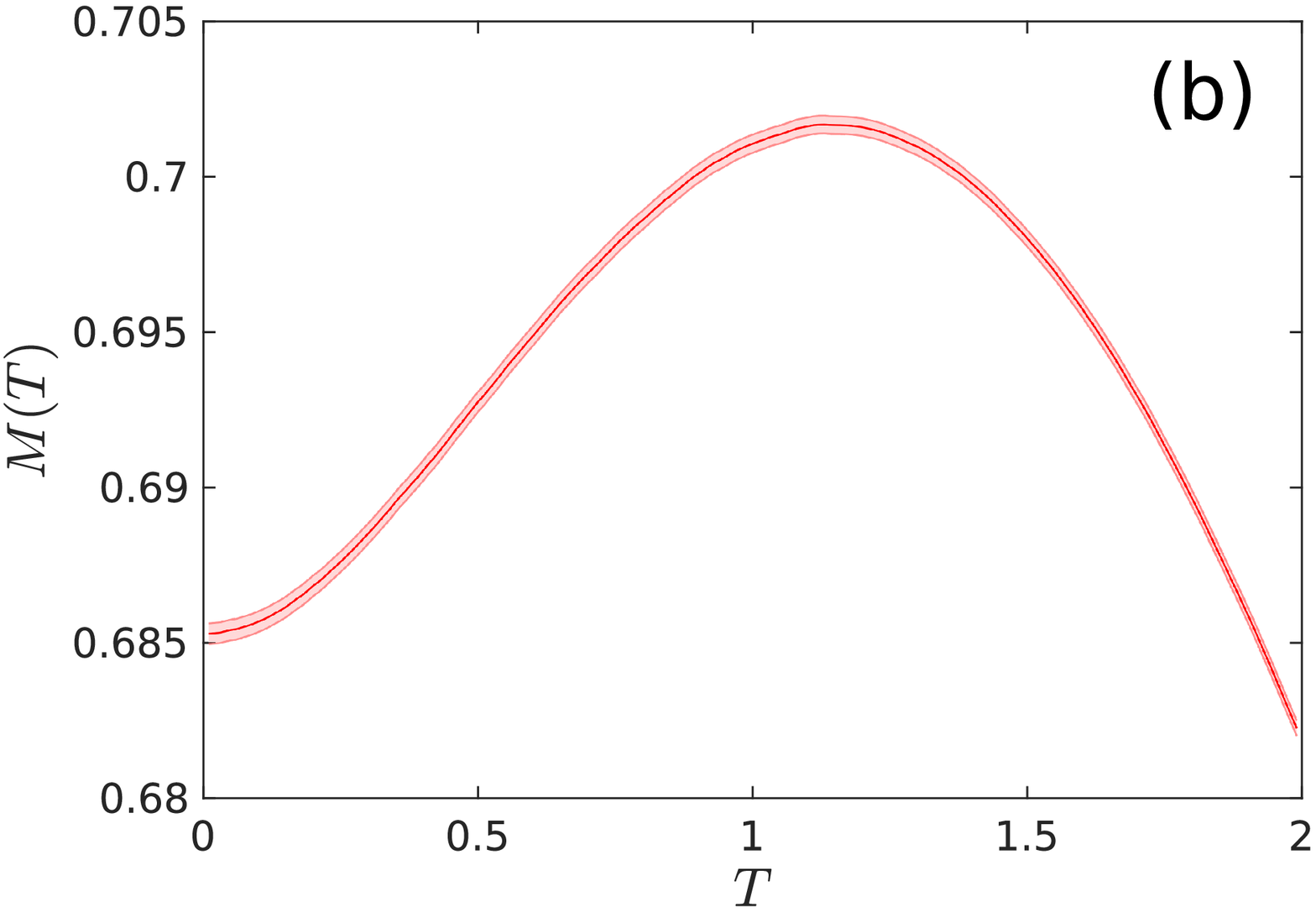}
		
			\includegraphics[width=\columnwidth]{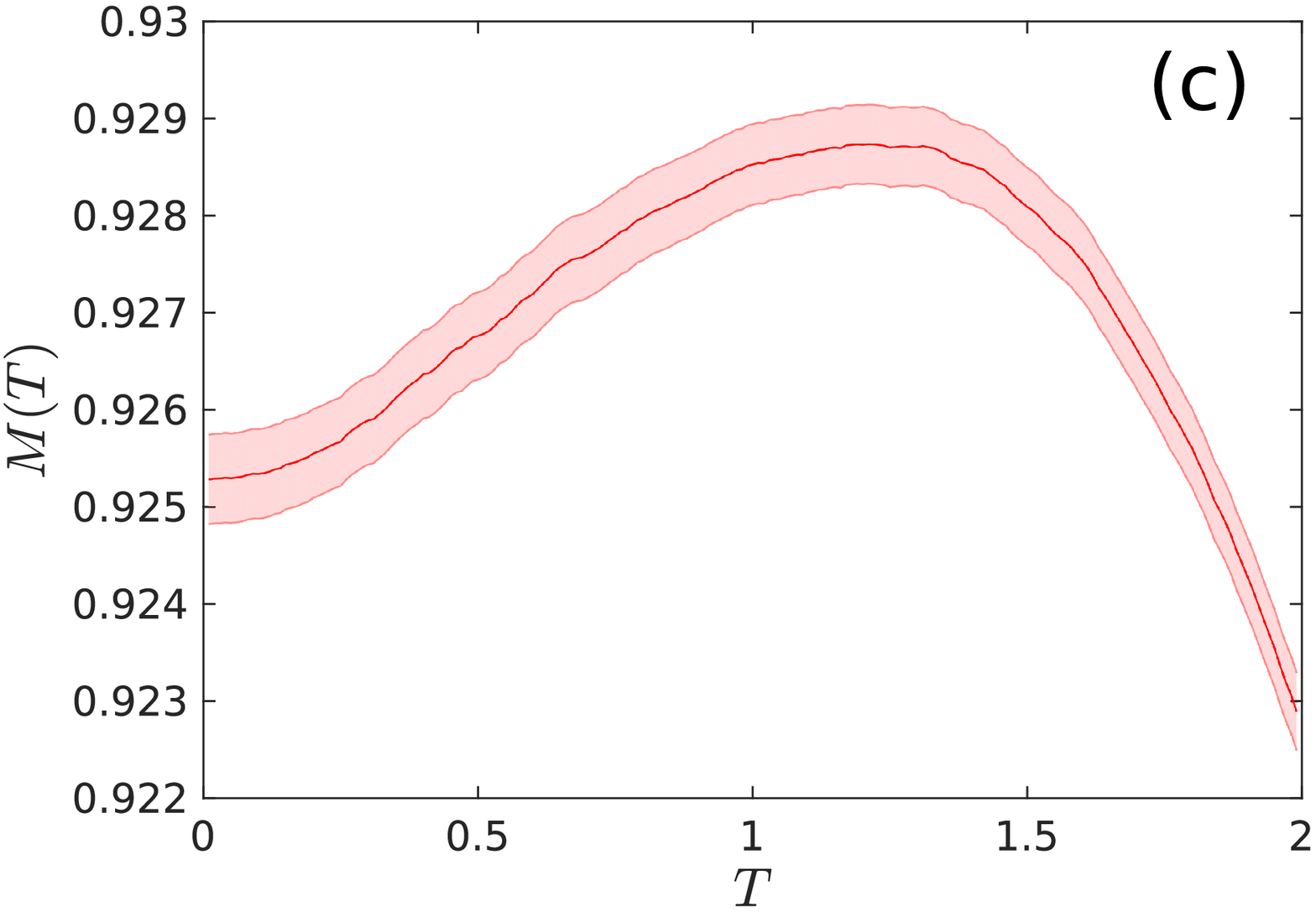}
			\includegraphics[width=\columnwidth]{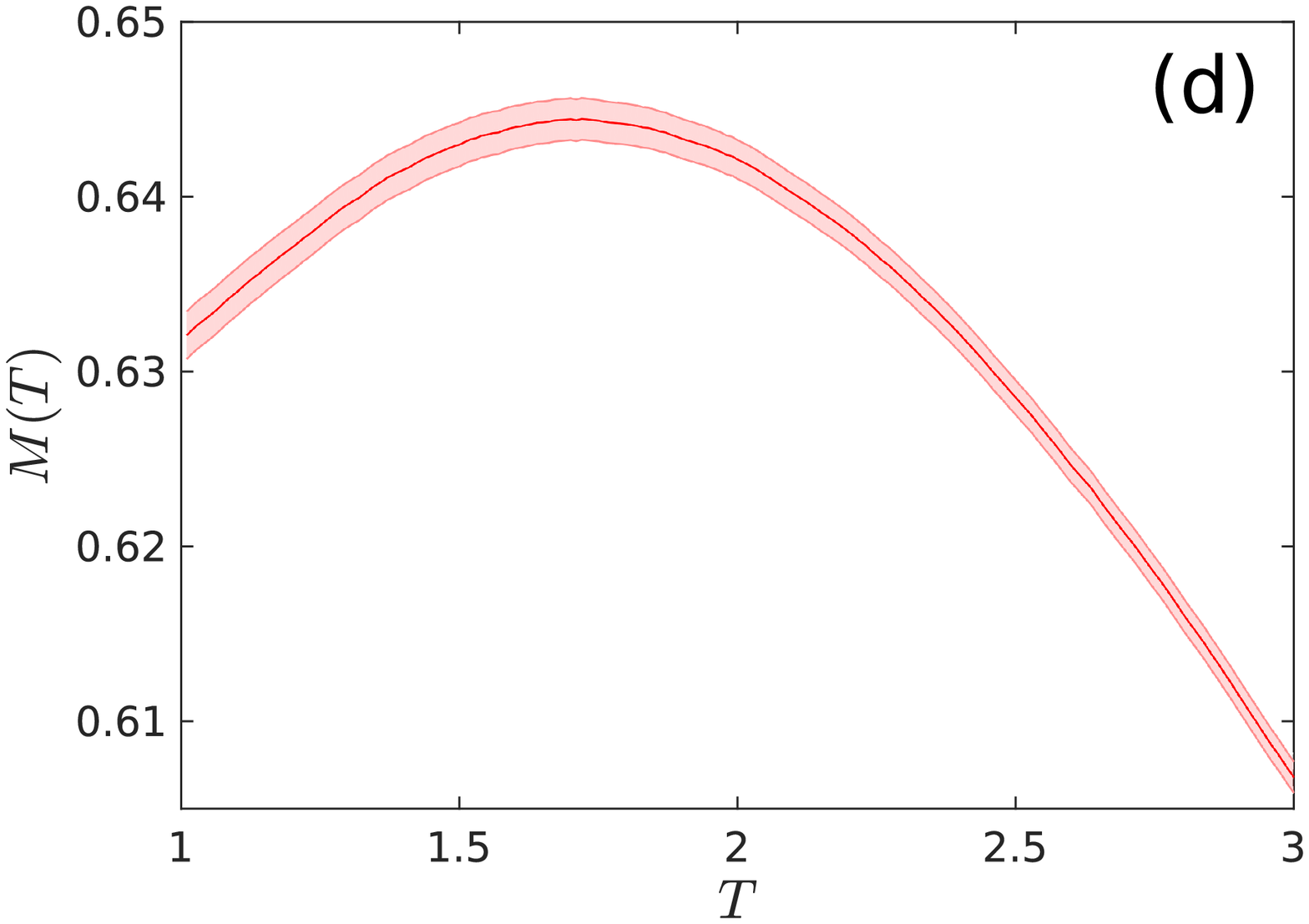}
			
			\caption{(Color online) Mean of the overlap $M(T)$ and its standard deviation as functions of the temperature $T$: data for the ladder (a) for $\sigma=1$ and $\gamma=0.5$ and (b) for $\sigma=2$ and $\gamma=0.4$; data for the stacked ladder (c) for $\sigma=1$ and $\gamma=0.5$ and (d) for $\sigma=2$ and $\gamma=0.4$.}
			\label{fig:2B}
\end{figure*}

Figure \ref{fig:2B} shows typical results for triangular and stacked ladders for several pairs of $\sigma$ and $\gamma$. The ordinate is the overlap $M(T)$ and the abscissa is the temperature $T$. The mean of the overlap is indicated by the red solid line, and the shaded area denotes the standard deviation. We observe a clear peak at finite temperature in each panel. Similar results were found for all pairs of $\sigma$ and $\gamma$. These results clearly show that the ground state of the original Hamiltonian can be inferred with higher probability from data generated by the noisy Hamiltonian at finite temperature than at zero temperature.

\begin{figure*}
	\includegraphics[width=\columnwidth]{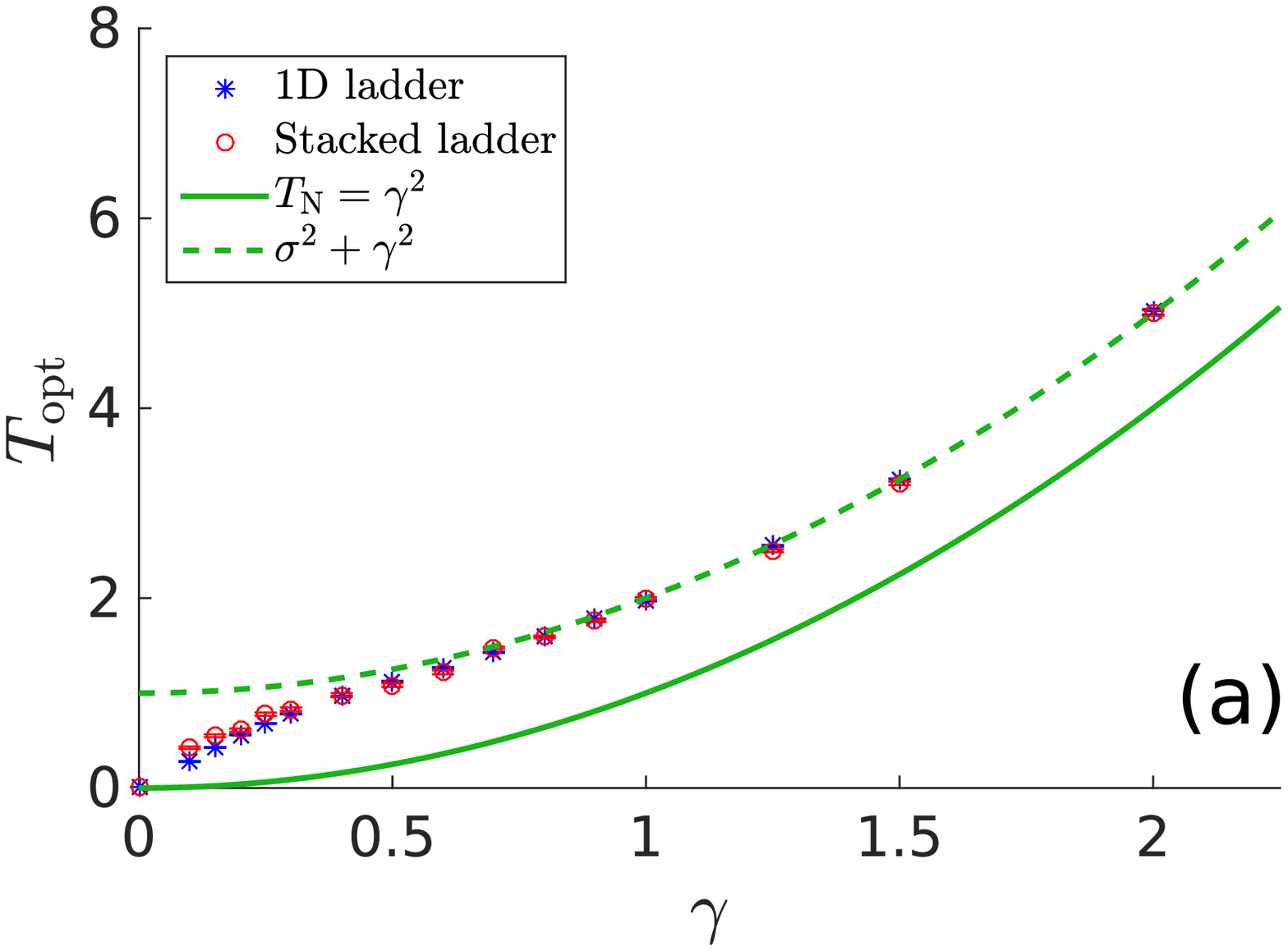}
	\includegraphics[width=\columnwidth]{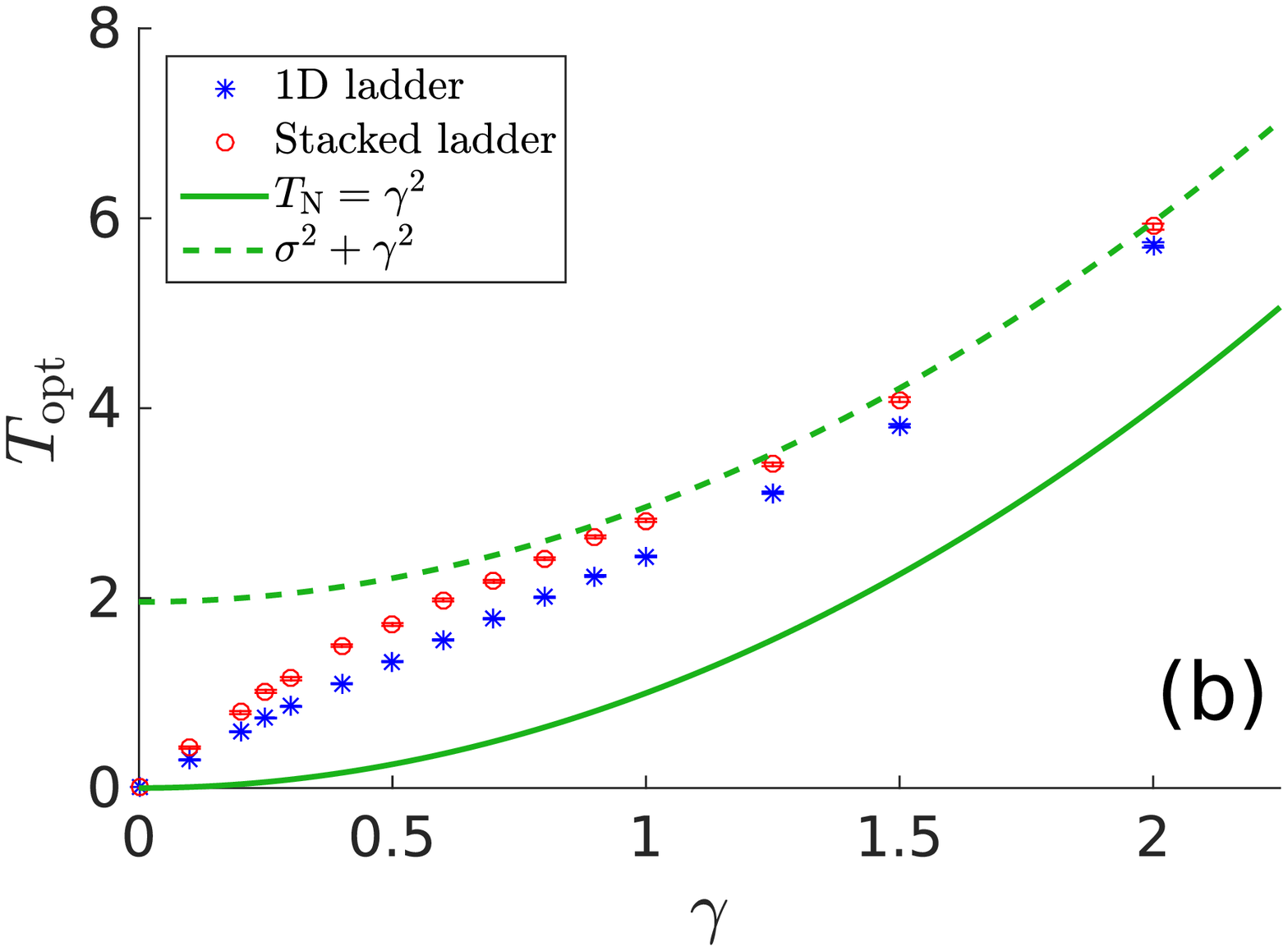}

	\includegraphics[width=\columnwidth]{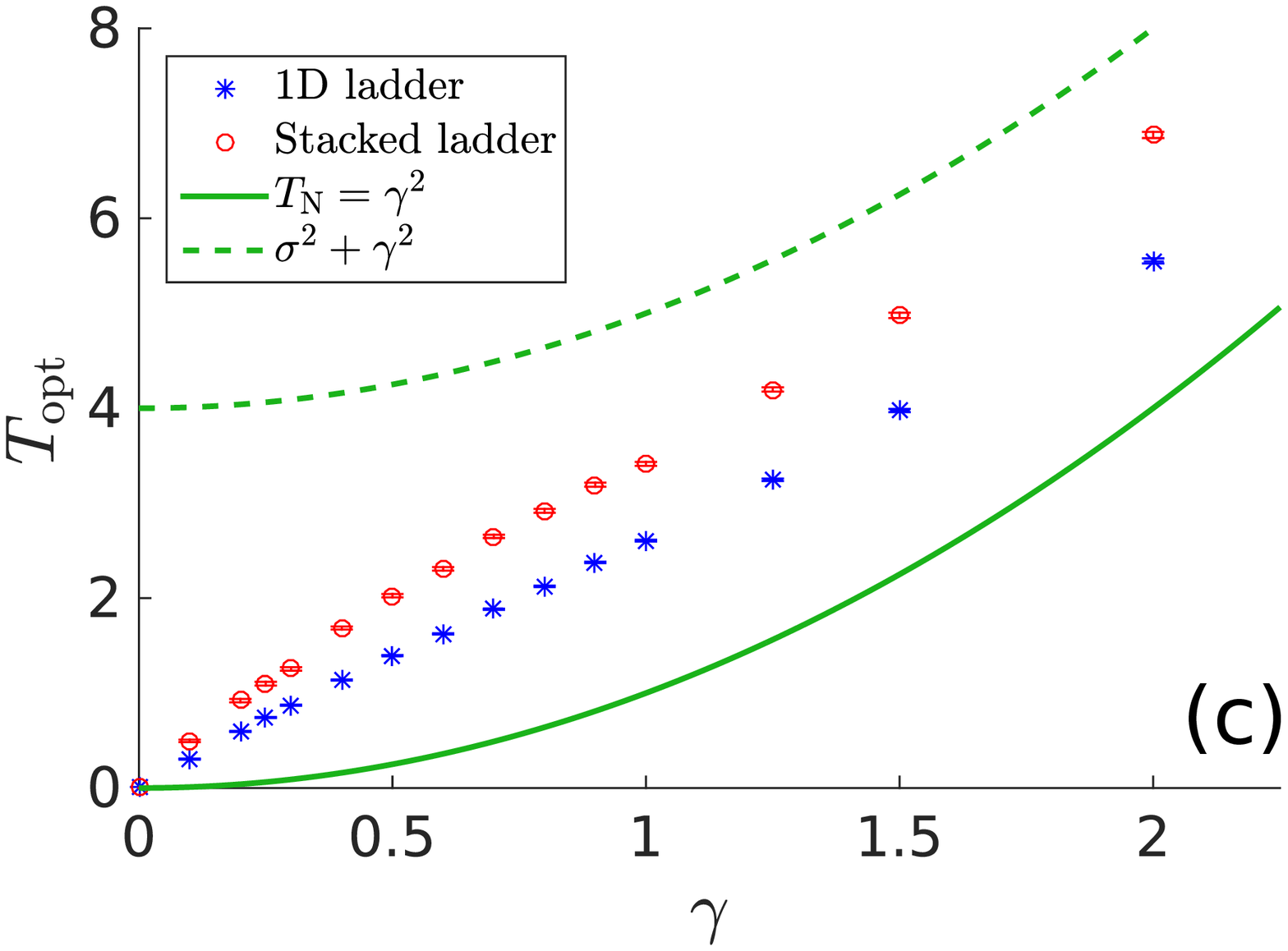}
	\includegraphics[width=\columnwidth]{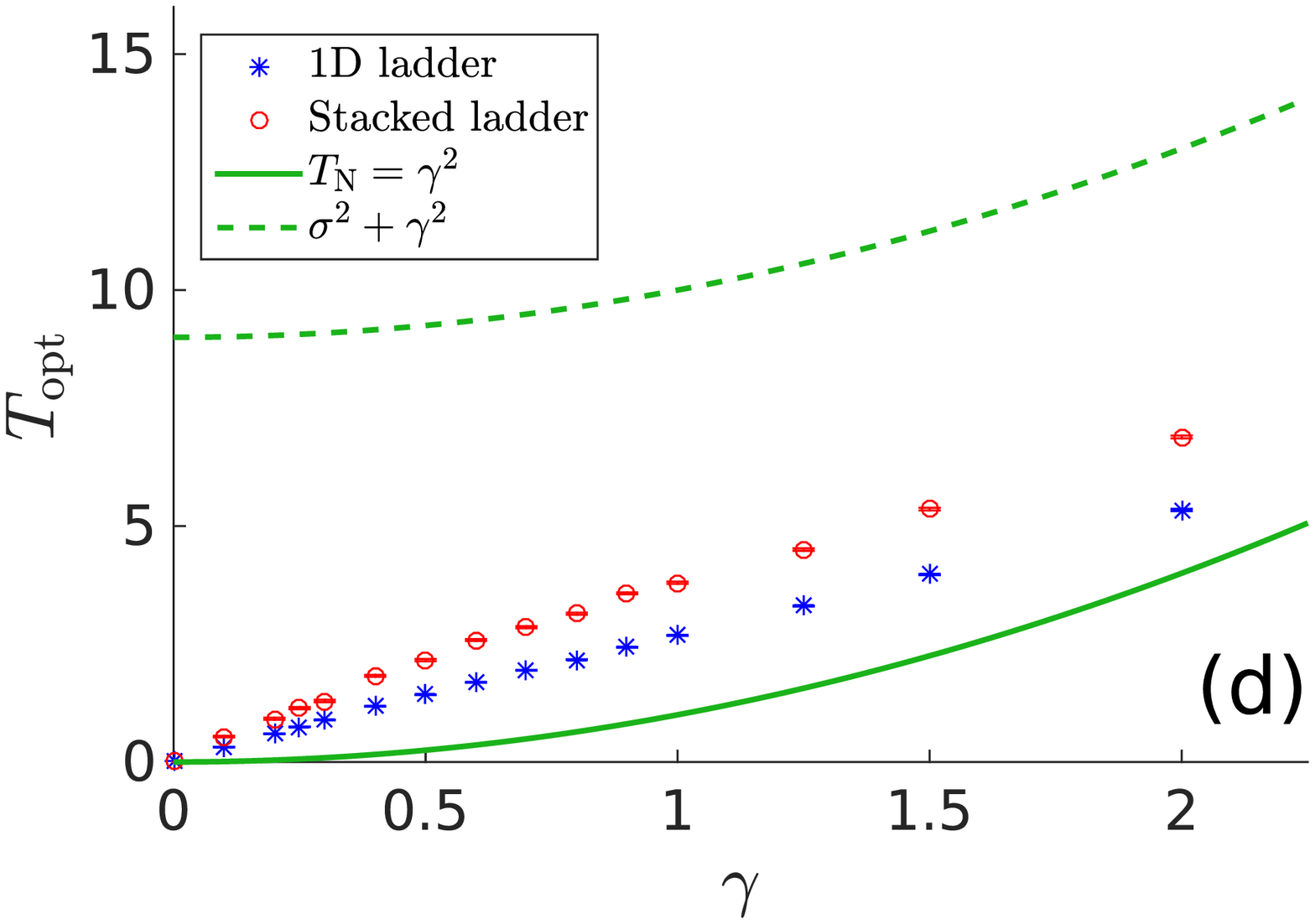}
	
	\caption{(Color online) Optimal decoding temperature $T_{\mathrm{opt}}$ as a function of the standard deviation of the noise $\gamma$. The green solid curve in each panel represents the optimal temperature for $\sigma = 0$ ($T_{\mathrm{N}} = \gamma^2$) and the green dashed line shows the result of the mean-field analysis ($\sigma^2+\gamma^2$). (a) $\sigma = 1$, (b) $\sigma = 1.4$, (c) $\sigma = 2$, and (d) $\sigma = 3$}.
	\label{fig:2B2}
\end{figure*}

Figure \ref{fig:2B2} shows the optimal temperature $T_{\mathrm{opt}}$ for decoding, i.e., the peak position of $M(T)$, as a function of $\gamma$ for each fixed value of $\sigma$. The green curve is the optimal temperature for $\sigma = 0$ ($T_{\mathrm{N}} = \gamma^2$). The data indicate that the naive relation $T_{\mathrm{opt}}  = \gamma^2$ for $\sigma=0$ does not hold when $\sigma >0$. The functional form of $T_{\mathrm{opt}}(\gamma)$ thus depends on the lattice structure, unlike the case of $\sigma = 0$.

\section{Mean-field analysis} 
\label{sec:MFA}

Because it is difficult to develop a generic theory to derive $T_{\rm opt}(\gamma)$ for the noisy case when $\sigma >0$, we use mean-field theory to understand its behavior qualitatively. The original Hamiltonian is chosen to be the Sherrington-Kirkpatrick (SK) fully connected Ising spin glass \cite{SKmodel},
\begin{align}
	\label{eq:3:origH}
	H_{\rm SK}(\v{S}) = - \sum_{i<j}J_{ij}S_i S_j ,
\end{align}
where the sum runs over all distinct pairs of spins. The interactions are drawn from a Gaussian distribution
\begin{align}
	P(J_{ij})=\frac{1}{\sigma}\sqrt{\frac{N}{2 \pi}} \exp{\left\{ -\frac{N}{2\sigma^2} \left( J_{ij}-\frac{J_0}{N}\right)^2\right\}},
\end{align}
where $J_0$ represents the mean of the distribution and $\sigma^2$ its variance. The Hamiltonian with noise is given by
\begin{align}\label{eq:3:noisyH}
	\tilde{H}_{\rm SK}(\v{\tau}) &= - \sum_{i<j}\tilde{J}_{ij}\tau_i \tau_j,\quad (\tau_i=\pm 1) , 
\end{align}
where the interactions are affected by noise via
\begin{align}
	\tilde{J}_{ij} = J_{ij}+\xi_{ij}.
\end{align}
As before, the noise follows a Gaussian distribution with variance $\gamma^2$,
\begin{align}
	P(\xi_{ij})=\frac{1}{\gamma}\sqrt{\frac{N}{2 \pi}} \exp{\left( -\frac{N}{2\gamma^2} \xi_{ij}^2\right)} .
\end{align}
The overlap is
\begin{align}
	M(T) = \lim_{\beta_0 \to \infty}[\mathrm{sgn}\braket{S_i}_{\beta_0}\mathrm{sgn}\braket{\tau_i}_{\beta}],
\end{align}
where the square brackets denote the configurational average over the distributions of interactions and noise,
\begin{align}
	\int\prod_{i<j}dJ_{ij}d\xi_{ij}P(J_{ij})P(\xi_{ij})(\cdots) \equiv [\cdots].
\end{align}
The angular brackets represent the thermal average with respect to each Hamiltonian:
\begin{align}
	\begin{split}
		\braket{S_i}_{\beta_0} &= \frac{\mathrm{Tr}_{{\scriptsize \v{S}}}\,S_{i}\exp{[-\beta_0 H(\v{S})]}}{\mathrm{Tr}_{{\scriptsize \v{S}}}\exp{[-\beta_0 H(\v{S})]}}, \\
		\braket{\tau_i}_{\beta} &= \frac{\mathrm{Tr}_{\v{\tau}}\,\tau_{i}\exp{[-\beta \tilde{H}(\v{\tau})]}}{\mathrm{Tr}_{\v{\tau}}\exp{[-\beta \tilde{H}(\v{\tau})]}}.
	\end{split}
\end{align}
We assume $\beta_0$ to be finite in the course of the calculations and take the limit $\beta_0 \to \infty$ at the end. Our goal is to identify the temperature that maximizes the overlap. As detailed in the Appendix, the free energy per spin is calculated under the ansatz of replica symmetry as
\begin{align}
	\begin{split}
	\label{eq:3:expfree}
		-[f] &= \frac{\sigma^2 \beta_0^2}{4}q_0^2+\frac{(\sigma^2+\gamma^2) \beta^2}{4}q^2-\frac{J_0 \beta_0}{2}{m_0}^2-\frac{J_0 \beta}{2} m^2 \\
		&+\frac{\sigma^2 \beta_0^2}{4}+\frac{(\sigma^2+\gamma^2)\beta^2}{4} 
		-\frac{\sigma^2\beta_0^2}{2}q_0-\frac{(\sigma^2+\gamma^2)\beta^2 q}{2}\\
					&+\int Dz_0 \ln (2\cosh H_1(z_0)) \\
					&+\int Dz \ln (2\cosh H_2(z)),
	\end{split}
\end{align}
where
\begin{align}
    H_1(z_0) &= \sqrt{\sigma^2\beta_0^2 q_0}\,z_0+J_0\beta_0 m_0,\\
    H_2(z_0) &= \sqrt{(\sigma^2+\gamma^2)\beta^2 q}\,z_0+J_0\beta m.
\end{align}
The self-consistent equations for the order parameters read
\begin{align}
	\begin{split}
	m_0 &= \int Dz_0 \tanh   H_1(z_0)=[\langle S_i\rangle_{\beta_0}],\\
	m   &= \int Dz   \tanh   H_2(z  )=[\langle \tau_i\rangle_{\beta}],\\
	q_0 &= \int Dz_0 \tanh^2 H_1(z_0)=[\langle S_i\rangle_{\beta_0}^2],\\
	q   &= \int Dz   \tanh^2 H_2(z  )=[\langle \tau_i\rangle_{\beta}^2].
	\end{split}
\end{align}

As shown in the Appendix, the two systems defined in Eqs.~(\ref{eq:3:origH}) and (\ref{eq:3:noisyH}) decouple in the replica symmetric solution, and we obtain
\begin{align}
	\begin{split}
	M(T) &= \lim_{\beta_0 \to \infty}[\mathrm{sgn}\braket{S_i}_{\beta_0}\mathrm{sgn}\braket{\tau_i}_{\beta}]  \\
	&= \lim_{\beta_0 \to \infty} \left[\mathrm{sgn}\Braket{S_i}_{\beta_0}\right]_{\sigma^2} \left[\mathrm{sgn}\Braket{\tau_i}_{\beta}\right]_{\sigma^2+\gamma^2},
	\end{split}
\end{align} 
where the subscript in the outer square brackets in the second line denotes the variance of the distribution of randomness. Since the first factor,
\begin{align}
  \lim_{\beta_0 \to \infty}[\mathrm{sgn}\braket{S_i}_{\beta_0}]_{\sigma^2},
\end{align}
is independent of $\beta$, only the second factor,
\begin{align}
  \left[\mathrm{sgn}\Braket{\tau_i}_{\beta}\right]_{\sigma^2+\gamma^2},   
\end{align}
determines the temperature dependence of $M(T)$. The latter is nothing but the well-established case of vanishing original variance \cite{rujan1993,nishimori1993,sourlas1994,iba1999,nsmrbook} because the expression involves a single system with noise. Therefore, we reach the conclusion that the overlap takes a maximum value at the temperature
\begin{align}
    T_{\rm opt} = \sigma^2+\gamma^2.
\end{align}
This result is naturally consistent with the case of vanishing original variance $\sigma = 0$, i.e., $T_{\mathrm{N}} = \gamma^2$.

The green dashed line depicted in Fig.~\ref{fig:2B2} shows the mean-field prediction $T_{\rm opt} = \sigma^2 + \gamma^2$. The numerical data agree relatively satisfactorily with the mean-field prediction for small $\sigma$ and large $\gamma$. The finite value of $T_{\rm opt}$ at $\gamma =0$ should be an artifact of the replica symmetric solution because the limit $\gamma =0$ represents the noiseless case, where the original and the noisy Hamiltonians coincide, and thus $T=0$ should give the best result.

%

\section{Conclusion} 
\label{sec:Conclusion}

We have examined the effects of noise added to the interactions of an  Ising spin-glass Hamiltonian. The goal is to infer the ground state of the original Ising Hamiltonian using only the output from the system with noisy interactions. The spin configuration of the latter system at finite temperature has been compared with the ground state of the original system. It has been shown from numerical transfer-matrix calculations that the overlap of two spin configurations has a maximum at a {\em finite} temperature, thus giving the smallest Hamming distance. This means that the ground state of the original Hamiltonian is better inferred at finite temperature than at the ground state of the noisy system. An intuitive explanation of this result would be that the unperturbed ground state is an excited state of the noisy system. Therefore, a finite temperature that corresponds to the excitation energy might increase the chances to find the ground state of the original problem Hamiltonian. Nevertheless, the original ground state is just one of very many states to be realized at a specific finite temperature for the noisy system and our result is quite nontrivial.

A similar phenomenon has been known to exist for years for the case of uniform ferromagnetic interactions in the original Hamiltonian under the context of error-correcting codes \cite{rujan1993,nishimori1993,sourlas1994,iba1999,nsmrbook,chancellor:16}.  The present work generalizes this old result to the case with randomness already in the original model. Although the former case of uniform original interactions has been able to be treated analytically with full generality in the sense that there is no restriction on the type of lattice or the range of interactions \cite{nsmrbook,nishimori1993,sourlas1994,iba1999}, it is difficult to develop a comparable analytical theory for the present case because of the difference in symmetries. We therefore used the numerical transfer-matrix method and a mean-field analysis. It is an important direction of further research to establish analytical results under general conditions for the case with random original interactions. In particular, it would be very useful to derive an explicit analytical expression for the optimal temperature that is applicable beyond mean-field theory.

Finiteness of the optimal temperature implies an important lesson for data analysis (postprocessing) of a real quantum annealer. Because any device operates at finite temperature, it is always a serious problem to keep temperature effects under control. However, our result suggests that thermal noise may be positively used to infer the ground state of the original Hamiltonian when the interactions are disturbed by noise. For example, if the original Hamiltonian has uniformly ferromagnetic interactions, the optimal temperature is given as
\begin{equation}
    T_{\rm N}=J_0 \left(\frac{\gamma}{J_0}\right)^2,
\end{equation}
where $J_0$ is the original uniform interaction and $\gamma^2$ is the variance of noise. Notice that we set $J_0=1$ in Sec.~\ref{sec:Formulation}. If we insert the values roughly corresponding to those for the D-Wave machine, $J_0=5$ GHz and $\gamma/J_0=0.1$, the latter representing the standard deviation of control errors, the optimal temperature turns out to be approximately $2$ mK.  The operating temperature of the D-Wave machine is approximately $20$ mK and, therefore, much higher than this theoretical optimal temperature. It should nevertheless be remarked that the numbers would easily change if the original interactions are random, if the real control errors are larger than $\gamma/J_0 = 0.1$, or if the noise is distributed according to a another distribution (e.g., $1/f$ or pink noise). It has indeed been shown in our numerical calculations that the optimal temperature for random original interactions is much higher than $T_{\rm N}$ as seen in Fig.~\ref{fig:2B2}. It is, however, noteworthy that our theoretical estimate is not much different than the real operating temperature of the device. Therefore, it would be interesting to verify these theoretical predictions on the actual D-Wave 2X device for nontrivial spin-glass problems complementing previous results on the trivial ferromagnetic case \cite{chancellor:16}.

We have chosen the overlap $M(T)$ as the measure of better inference of the ground state. This means to minimize the Hamming distance to the true ground state, which is the standard in error-correcting codes \cite{nsmrbook}.  This measure is also useful in other circumstances including the spin-glass problem, where the spin configuration of the ground state is of as much interest as the ground-state energy.  In the context of combinatorial optimization problems, however, it is often the case that the energy is a prime measure of performance. It is useful to remember here that a smaller Hamming distance to the ground state does not necessarily mean a smaller energy \cite{pudenz:15}. It may then happen that our present result does not apply as is, if it is desirable to minimize the energy, not to minimize the Hamming distance.  This turns out to be a highly nontrivial problem and will be discussed in a forthcoming paper. It is also interesting to see how our conclusion would change if quantum effects are taken into account explicitly.  We are studying this problem and results will be announced before too long.

\begin{acknowledgments}

The work of H.N.~was funded by the ImPACT Program of Council for Science, Technology and Innovation, Cabinet Office, Government of Japan, and by the JPSJ KAKENHI Grant No. 26287086. A.J.O.~and H.G.K.~acknowledge support from the National Science Foundation (Grant No.~DMR-1151387). The work of H.G.K.~and A.J.O.~is supported in part by the Office of the Director of National Intelligence (ODNI), Intelligence Advanced Research Projects Activity (IARPA), via MIT Lincoln Laboratory Air Force Contract No.~FA8721-05-C-0002. The views and conclusions contained herein are those of the authors and should not be interpreted as necessarily representing the official policies or endorsements, either expressed or implied, of ODNI, IARPA, or the U.S.~Government. The U.S.~Government is authorized to reproduce and distribute reprints for Governmental purpose notwithstanding any copyright annotation thereon.  We thank Texas A\&M University for access to their HPC resources.

\end{acknowledgments}

\appendix
\section{Replica symmetric solution}\label{section:app1}

In this Appendix, we explain the derivation of the replica-symmetric solution of the mean-field model discussed in Sec.~\ref{sec:MFA}. The calculation follows the standard {\em modus operandi}, but with two independent systems sharing part of the random interactions.

The configurational average of the $n$th power of the partition function is 
\begin{widetext}
	\begin{align}
		\begin{split}
			[Z^n] &= \int\prod_{i<j}dJ_{ij}d\xi_{ij}P(J_{ij})P(\xi_{ij})\mathrm{Tr}\exp\left\{\sum_{i<j}\sum_{\alpha=1}^{n}(\beta_0 J_{ij}S_{i}^{\alpha}S_{j}^{\alpha}+\beta(J_{ij}+\xi_{ij})\tau_{i}^{\alpha}\tau_{j}^{\alpha})\right\} \\
						&= \mathrm{Tr}\int\left(\prod_{i<j}dJ_{ij}d\xi_{ij}\right)\exp\left\{\sum_{i<j}\left(\beta_0 J_{ij}\sum_{\alpha=1}^{n}S_i^{\alpha}S_j^{\alpha}+\beta(J_{ij}+\xi_{ij})\sum_{\alpha=1}^{n}\tau_i^{\alpha}\tau_j^{\alpha}\right. \right. \\
					 &\left. \left. -\frac{N}{2\sigma^2}\left(J_{ij}-\frac{J_0}{N}\right)^2-\frac{N}{2\gamma^2}\xi_{ij}^2+\ln\left(\frac{1}{\sigma\gamma}\frac{N}{2\pi}\right)\right)\right\}.
		\end{split}
	\end{align}
The standard mean-field calculation \cite{nsmrbook} leads to 
\begin{align}
	\label{eq:3:fcomplete}
	\begin{split}
		-[f] &= \lim_{n \to 0}\left(-\frac{\sigma^2 \beta_0^2}{4n}\sum_{\alpha \neq \beta}{q_{\alpha \beta}^0}^2-\frac{(\sigma^2+\gamma^2) \beta^2}{4n}\sum_{\alpha \neq \beta}{q_{\alpha \beta}}^2-\frac{\sigma^2 \beta_0 \beta}{2n}\sum_{\alpha \beta}{u_{\alpha \beta}}^2 \right. \\
			&\left. -\frac{J_0 \beta_0}{2n}\sum_{\alpha}{m_{\alpha}^0}^2-\frac{J_0 \beta}{2n}\sum_{\alpha}{m_{\alpha}}^2+\frac{\sigma^2 \beta_0^2}{4}+\frac{(\sigma^2+\gamma^2)\beta^2}{4}+\frac{1}{n}\l{\mathrm{Tr}e^{L}}\right),
	\end{split}
\end{align}
where
\begin{align}
	\label{eq:3:L}
	\begin{split}
		L &\equiv \sigma^2 \beta_0^2\sum_{\alpha<\beta}q_{\alpha \beta}^0 S^{\alpha}S^{\beta}+(\sigma^2+\gamma^2) \beta^2\sum_{\alpha<\beta}q_{\alpha \beta} \tau^{\alpha}\tau^{\beta}+\sigma^2\beta_0 \beta \sum_{\alpha \beta}u_{\alpha \beta}S^{\alpha}\tau^{\beta} \\
		&+J_0 \beta_0 \sum_{\alpha}m_{\alpha}^0 S^{\alpha}+J_0 \beta \sum_{\alpha}m_{\alpha} \tau^{\alpha}.
	\end{split}
\end{align}
\end{widetext}
The order parameters in $[f]$ are determined by the saddle-point conditions
\begin{align}
	\begin{cases}
		q_{\alpha \beta}^0 &= \Braket{S^\alpha S^\beta}_L = \left[\braket{S_i^\alpha}_{\beta_0} \braket{S_i^\beta}_{\beta_0}\right] \\
		q_{\alpha \beta}   &= \Braket{\tau^\alpha \tau^\beta}_L = \left[\braket{\tau_i^\alpha}_{\beta}\braket{\tau_i^\beta}_{\beta}\right] \\
		m_{\alpha      }^0 &= \Braket{S^\alpha}_L = \left[\braket{S_i^\alpha}_{\beta_0}\right] \\
		m_{\alpha      }   &= \Braket{\tau^\alpha}_L = \left[\braket{\tau_i^\alpha}_{\beta}\right] \\
		u_{\alpha \beta}   &= \Braket{S^\alpha \tau^\beta}_L = \left[\braket{S_i^\alpha}_{\beta_0}\braket{\tau_i^\beta}_{\beta}\right],
	\end{cases}
\end{align}
where
\begin{align}
	\Braket{\cdots}_L \equiv \frac{\mathrm{Tr}(\cdots)e^L}{\mathrm{Tr}e^L}.
\end{align}

If we assume the replica-symmetric ansatz \cite{nsmrbook}, this free energy can be rewritten as follows:
\begin{align}
	\begin{split}
		-[f] &= \lim_{n \to 0}\left(-\frac{\sigma^2 \beta_0^2}{4n}(n^2-n)q_0^2-\frac{(\sigma^2+\gamma^2) \beta^2}{4n}(n^2-n)q^2\right. \\
		&\left. -\frac{\sigma^2 \beta_0 \beta}{2n}n^2u^2-\frac{J_0 \beta_0}{2n}n{m_0}^2-\frac{J_0 \beta}{2n}n m^2 \right. \\
			&\left. +\frac{\sigma^2 \beta_0^2}{4}+\frac{(\sigma^2+\gamma^2)\beta^2}{4}+\frac{1}{n}\ln{\mathrm{Tr}e^{L_0}}\right),
	\end{split}
\end{align}
where $L_0$ represents $L$ under the replica symmetric ansatz. The explicit form of $\ln{\mathrm{Tr}e^{L_0}}$ is \cite{comment:fn}
\begin{widetext}
\begin{align}
	\begin{split}
		\ln{\mathrm{Tr}e^{L_0}} &= \ln{\mathrm{Tr}\exp{\left\{\sigma^2 \beta_0^2 q_0 \sum_{\alpha<\beta}S^\alpha S^\beta +(\sigma^2+\gamma^2)\beta^2 q\sum_{\alpha<\beta}\tau^\alpha \tau^\beta \right.}} \\
		&\left.+\sigma^2 \beta_0 \beta u \left(\sum_{\alpha}S^\alpha\right)\left(\sum_{\beta}\tau^\beta\right)+J_0 \beta_0 m_0 \sum_{\alpha}S^\alpha+J_0 \beta m \sum_{\alpha}\tau^\alpha \right\} \\
		&=n\left(\int Dz_0 Dz Dw \ln \left[2\cosh H_1(z_0, w)\cdot 2\cosh H_2(z, w)\right]-\tilde{f}\right)+\mathcal{O}(n^2),
	\end{split}
\end{align}
\end{widetext}
where
\begin{align}\label{eq:H1H2}
	\begin{split}
	H_1(z_0, w) &= \sqrt{a}z_0+\sqrt{c}w+d, \\
	H_2(z, w) &= \sqrt{b}z+\sqrt{c}w+e, \\
	\end{split}
\end{align}
\clearpage
and
\begin{align}
\begin{cases}
	a &= \sigma^2\beta_0^2q_0-\sigma^2\beta_0 \beta u \\
	b &= (\sigma^2+\gamma^2)\beta^2 q-\sigma^2\beta_0 \beta u \\
	c &= \sigma^2\beta_0\beta u \\
	d &= J_0\beta_0 m_0 \\
	e &= J_0 \beta m \\
	\tilde{f} &= \frac{\sigma^2\beta_0^2q_0}{2}+\frac{(\sigma^2+\gamma^2)\beta^2 q}{2} \\
\end{cases}.\label{eq:abcdef}
\end{align}
Therefore, the free energy after taking the limit $n \to 0$ is
\begin{align}
	\begin{split}
		-[f] &= \frac{\sigma^2 \beta_0^2}{4}q_0^2+\frac{(\sigma^2+\gamma^2) \beta^2}{4}q^2-\frac{J_0 \beta_0}{2}{m_0}^2-\frac{J_0 \beta}{2} m^2 \\
		&+\frac{\sigma^2 \beta_0^2}{4}+\frac{(\sigma^2+\gamma^2)\beta^2}{4} \\
					&+\int Dz_0 Dw \ln (2\cosh H_1(z_0, w)) \\
					&+\int Dz Dw \ln (2\cosh H_2(z, w))-\tilde{f}.
	\end{split}
\end{align}
This result reduces to Eq.~(\ref{eq:3:expfree}) because the $w$ dependence of $H_1$ and $H_2$ disappears as explained below.  It turns out the averaged free energy does not depend on $u$, as can be verified by direct computations:
\begin{align}
	\label{eq:3:pdu0}
	-\pd{[f]}{u} = 0.
\end{align}
We can therefore choose $u=0$ without loss of generality. This implies that the original and noisy systems decouple completely because $u$ is the parameter that connects these two systems as seen in Eq.~(\ref{eq:3:L}).  When $u=0$, $c=0$ according to Eq.~(\ref{eq:abcdef}) and hence the $w$ dependence of $H_1(z_0,w)$ and $H_2(z,w)$ disappears as can be seen in Eq.~(\ref{eq:H1H2}).
\bibliographystyle{apsrev4-1}
\bibliography{reference.bib,reference_tamu.bib,comments.bib}

\begin{thebibliography}{38}
\bibitem [{com()}]{comment:fn}%
  \BibitemOpen
  \href@noop {} {}\bibinfo {note} {Note that $Dm \equiv
  \exp(-m^2/2)/\sqrt{2\pi} dm$ represents the Gaussian measure.}\BibitemShut
  {Stop}%
\end{thebibliography}

\end{document}